\documentclass[11pt]{article}
\usepackage{epsfig}
\def\beq{\begin{equation}}
\def\eeq{\end{equation}}
\def\bea{\begin{eqnarray}}
\def\eea{\end{eqnarray}}
\def\L{{\cal L}}
\def\p{\partial}
\def\dt{{d \over dt}~}
\def\vev#1{\left\langle #1 \right\rangle}
\def\x{{\lambda \over (4 \pi)^2}}
\begin{document}
\rightline{KOBE-TH-99-02}
\rightline{hep-ph/9902209}

\vspace{.5cm}

\begin{center}
{\LARGE Bose condensation at high temperatures}

\vspace{.5cm} 

Hidenori SONODA\footnote{E-mail: sonoda@phys.sci.kobe-u.ac.jp}\\ 
Physics Department, Kobe University, Kobe 657-8501, Japan

\vspace{.2cm} 
February 1999

\vspace{.2cm}
\end{center}

\begin{abstract}
Bose condensation is usually a low temperature phenomenon due to a low
particle number density.  When the number density is kept large compared
to the inverse Compton volume, Bose condensation can occur at a
temperature much higher than the mass of the particle.  We can then use
a three dimensional effective theory to study the thermal properties.
We compute the transition temperature for a complex scalar field theory
with a small interaction parameter.
\end{abstract}

\vskip .3cm

In a previous letter \cite{highT} we used an effective three dimensional
field theory \cite{Ginsparg, AP} to compute the critical temperature of
the weak coupling $\phi^4$ theory.  The purpose of this letter is to
extend the same technique to study theories with a chemical potential
coupled to a conserved particle number.  We will compute the transition
temperature $T_c$ for Bose condensation and the temperature dependence
of the specific heat near $T_c$.

It is the low number density of particles which makes Bose condensation
a low temperature phenomenon.  If the number density is large compared
to the inverse Compton volume $m^3$, where $m$ is the particle mass, the
transition temperature $T_c$ for Bose condensation becomes much higher
than $m$ so that $T_c$ is calculable using a three dimensional effective
theory.

In order to see this possibility more quantitatively, we consider a free
theory of spinless particles and antiparticles of mass $m$.  The
particle number density $n$ (or more precisely the number density of
particles minus that of antiparticles) is an increasing function of the
chemical potential $\mu$.  At the maximum chemical potential $\mu = m$,
we obtain the maximum number density without Bose condensation.  For
temperature $T$ much higher than $m$, the maximum number density is
approximately
\beq
n \simeq {m T^2 \over 2} + {4 m^3 \over 3 (4 \pi)^2}~.\label{nfree}
\eeq
For $n$ larger than the right-hand side, Bose condensation takes place.
Hence, for a given $n$, the transition temperature $T_c$ is given by
\beq
T_c^2 \simeq {2 \over m} \left( n - {4 m^3 \over 3
(4\pi)^2}\right)~.\label{Tcfree}
\eeq
This implies $T_c \gg m$ for $n \gg m^3$.  The above result
(\ref{Tcfree}) is expected to be valid with a small correction in the
presence of weak interactions among the bosons.

For concreteness we consider a complex scalar field theory with a global
U(1) symmetry in thermal equilibrium at temperature $T$.  The lagrangian
density in the four dimensional euclidean space is given by
\beq
\L_0 = \p_\mu \phi^* \p_\mu \phi + m^2 \phi^* \phi +
{\lambda \over 4} \left(\phi^* \phi\right)^2 + (\rm{counterterms})~,
\eeq 
where 
$\phi$ is a complex scalar field which is periodic in the
euclidean time direction:
\beq
\phi (\vec{x},\tau + 1/T) = \phi (\vec{x}, \tau)~.
\eeq
The counterterms are given in the $\overline{\rm MS}$ scheme.  Hence,
the renormalized parameters satisfy the renormalization group (RG)
equations
\bea
\dt \x &=& - 5 \left(\x\right)^2 + 15 \left( \x \right)^3 + ...~, \\
\dt m^2 &=& \left( 2 - 2 ~\x + {5 \over 2} \left(\x \right)^2 +
... \right) m^2~.
\eea
We can introduce the chemical potential $\mu$ as the euclidean time
component of an external U(1) gauge field:
\beq
A_\tau = i \mu~.
\eeq
Note this is purely imaginary.  The Ward identity protects it from
renormalization.  The total lagrangian density is therefore
\beq 
\L = \L_0 - \mu Z~\phi^* \stackrel{\leftrightarrow}{\p_\tau}
\phi - \mu^2 Z \phi^* \phi~,\label{lfour}
\eeq
where $Z$ is the wave function renormalization constant.  The average
number density of particles, $n$, is the first order derivative of the
free energy density $Y_4 (T,\mu)$ with respect to $\mu$:
\beq
n = - \left({\p Y_4 \over \p \mu}\right)_T~. \label{number}
\eeq
Our task is to find the transition temperature $T_c$ so that for $T <
T_c$ the field $\phi$ gets a non-vanishing expectation value.  We will
find $T_c$ first for a given chemical potential $\mu$, and then for a
given number density $n$.

A naive loop expansion of $Y_4$ suffers from infrared divergences at
two-loop and beyond \cite{Shaposhnikov}.  This difficulty is best
avoided by reducing the theory to a three dimensional effective theory
\cite{Ginsparg, AP} whose infrared properties are much better understood
\cite{FKRS, BN}.  The effective theory is given by the following
lagrangian density:
\beq
\L_3 = g_3 + \p_\mu \phi_3^* \p_\mu \phi_3 + m_3^2 ~\phi_3^* \phi_3 +
{\lambda_3 \over 4} (\phi_3^* \phi_3)^2 + ({\rm counterterms})~,
\eeq
where the counterterms are given in the $\overline{\rm MS}$ scheme.  The
higher dimensional interaction terms are negligible within our
approximation. The RG equations of the parameters are given by
\beq
\dt g_3 = 3 g_3 + \tilde{A} \lambda_3^3~,\quad
\dt m_3^2 = 2 m_3^2 + \tilde{C} \lambda_3^2~,\quad
\dt \lambda_3 = \lambda_3~,\label{RGthree}
\eeq
where
\beq
\tilde{A} = {1 \over (4 \pi)^2} {5 \over 3 \cdot 2^9}~,\quad
\tilde{C} = - {1 \over (4 \pi)^2} {1 \over 2}~.\label{AC}
\eeq
The parameters of the two theories are related such that the free energy
density $F_3 (g_3, m_3^2, \lambda_3)$ of the effective theory reproduces
that of the original theory:
\beq
Y_4 (T, \mu) = T F_3 (g_3, m_3^2, \lambda_3)~.\label{equivalence}
\eeq

For the three dimensional reduction to be valid, the temperature must be
high compared to $m$, $\mu$:
\beq
T \gg m ~{\approx}~ \mu~.
\eeq
Then, we can choose a renormalization scale $\Lambda$ such that
\beq
T/\Lambda = {\rm O} (N)~,\quad
m^2 /\Lambda^2 = {\rm O} (N)~, \quad
\mu^2 /\Lambda^2 = {\rm O} (N)~, \label{large}
\eeq
where O($N$) denotes an order of $N$, a large number.  We consider such
a range of $T$ so that we can identify the smallness of $1/N$ with the
smallness of the coupling $\lambda$:
\beq
\lambda = {\rm O} (1/N)~.\label{small}
\eeq
Eqns.~(\ref{large},\ref{small}) guarantee that we can calculate the
parameters of the effective theory in powers of $\lambda$, $m^2/T^2$,
and $\mu^2/T^2$ which are all of order $1/N$.  Since the calculation is
straightforward, we omit the detail and only state the result:
\bea
\lambda_3 &\simeq& \lambda T\label{lambda3}~,\\
m_3^2 &\simeq& m^2 - \mu^2 + {\lambda \over 12}~T^2 
+ \x \left[ 2 m^2 (\ln T/\Lambda + j_2) - 2
\mu^2 \right]\nonumber\\
&& + {\lambda^2 \over (4 \pi)^2}~T^2 \left[ - {1 \over 12}
\ln T/\Lambda + {j_2 \over 6} - {j_3 \over 4}\right]~, \label{m3}\\
T g_3 &\simeq& T^4 {(4\pi)^2 \over 144} \left[ - {1 \over 5} + {1 \over
2} \x + \left(\x\right)^2 \left( {5 \over 2} \ln T/\Lambda + j_2 + {3
\over 2} j_4\right) \right] \nonumber\\
&& + {T^2 \over 12} \left[ m^2 \left( 1 + 2 \x (\ln T/\Lambda + j_2)
\right) + \mu^2 \left( - 3 - 2 \x \right) \right] \nonumber \\
&& {}+ {1 \over (4 \pi)^2} \left[ m^4 (\ln T/\Lambda + j_2) + {2 \over 3}
\mu^4 - 2 m^2 \mu^2 \right]~,\label{g3}
\eea
where the constants are given by\footnote{The necessary integrals have
been calculated by Arnold and Zhai \cite{AZ}.}
\bea
j_2 &=& \ln 4 \pi - \gamma~,\quad
j_3 = \ln 4 \pi -1 - {\zeta' (-1) \over \zeta (-1)}~,\nonumber\\
j_4 &=& \ln 4 \pi - {31 \over 30} - 2 {\zeta' (-1) \over \zeta (-1)} +
{\zeta'(-3) \over \zeta (-3)}~.\label{constants}
\eea
In the above we have computed $\lambda_3, m_3^2/\Lambda^2$ to order
$N^0$ and $T g_3/\Lambda^4$ to order $N^2$.

Let us determine the transition temperature $T_c$ as a function of the
chemical potential $\mu$.  In the effective theory the expectation value
$\vev{\phi_3} \propto \vev{\phi}$ is non-vanishing if
\beq
R (m_3^2, \lambda_3) < R_c~,
\eeq
where $R(m_3^2, \lambda_3)$ is an RG invariant defined by
\cite{highT,Parisi}
\beq
R (m_3^2, \lambda_3) \equiv {m_3^2 \over \lambda_3^2}
- \tilde{C} \ln \lambda_3~. \label{R}
\eeq
Only a non-perturbative calculation can determine $R_c$, and we must
leave it as an unknown constant here.  By substituting
eqns.~(\ref{lambda3},\ref{m3}) into $R=R_c$, we obtain the transition
temperature \cite{highT}:
\bea
\lefteqn{{\lambda \over 12} ~T_c^2 \simeq \mu^2-m^2} \nonumber\\ 
&& + \x \Bigg[ 12 (\mu^2-m^2) \left( (4 \pi)^2 R_c - {1 \over 2} 
\ln \lambda - {5 \over 24} \ln {12 (\mu^2 - m^2) \over \lambda
\Lambda^2} + {j_3 \over
4} - {j_2 \over 6} \right)\nonumber\\ && \hspace{1.5cm} - m^2 \ln {12
(\mu^2 - m^2) \over \lambda \Lambda^2} - 2 m^2 j_2 + 2 \mu^2 \Bigg]~,
\label{Tc}
\eea
where the constants $j_2, j_3$ are given by eqns.~(\ref{constants}).  The
above gives $T_c/\Lambda$ to order $N^0$.  The dependence on $\ln
\lambda$ is the source of infrared divergences in the naive loop
expansions.

Eqn.~(\ref{Tc}) gives $T_c$ as a function of the chemical potential
$\mu$, but it is more convenient to express $T_c$ as a function of the
number density $n$.  To do this, we must express $\mu$ as a function of
$n$ by inverting eqn.~(\ref{number}).  The equivalence
(\ref{equivalence}) implies that we must obtain the free energy density
$F_3$ of the effective theory.  Since the cosmological constant $g_3$
has nothing to do with interactions, we find
\beq
F_3 (g_3, m_3^2, \lambda_3) = g_3 + f_3 (m_3^2, \lambda_3)~.\label{F3}
\eeq
The RG eqns.~(\ref{RGthree}) imply that the function $f_3$ can be
written as
\beq
f_3 (m_3^2, \lambda_3) = \lambda_3^3 \left( - \tilde{A} \ln \lambda_3 +
\overline{f_3} (R) \right)~,\label{f3}
\eeq
where $\tilde{A}$ is given in eqns.~(\ref{AC}), and $\overline{f_3}$ is
a function of the RG invariant $R$ (\ref{R}).  The theory of critical
phenomena gives the following scaling formula near $R=R_c$
\cite{Parisi,exponents}:
\beq
\overline{f_3} (R) = \overline{f_3} (0) + a ~|R-R_c|^{3 \over y_E} +
{\rm O} \left( |R-R_c|^{3-y' \over y_E}\right)~,\label{f3bar}
\eeq
where $a >0$ is a constant.  The constants $y_E$ and $y'$ are the
critical exponents of the three dimensional XY model: $y_E >0$ is the
scale dimension of the relevant parameter, and $y'<0$ is that of the
least irrelevant parameter.  They are given approximately as
\cite{exponents}
\beq
y_E \simeq 1.6~,\quad y' \simeq - 0.4~.
\eeq
Substituting the above results into eqn.~(\ref{number}), we obtain the
relation between $n$ and $\mu$
\beq
n = 2 \mu T^2 \left[ {1 \over 4} - {1 \over (4\pi)^2} {{4 \over 3} \mu^2
- 2 m^2 \over T^2} + \x \left( {1 \over 6} + (4\pi)^2 \overline{f_3}'
(R) \right) \right] \label{nvsmu}~
\eeq
up to terms of order $N^{3 \over 2} \Lambda^3$.  Note that the
derivative $\overline{f_3}'(R)$ vanishes at criticality $R=0$.

Solving eqns.~(\ref{Tc},\ref{nvsmu}), we can obtain the critical
temperature $T_c$ as a function of the number density $n$ to the order
$N^0 \Lambda$.  For simplicity, however, we will only present the result
at the leading order which is $N \Lambda$.  Eqns.~(\ref{Tc},\ref{nvsmu})
give
\beq
{1 \over 12} \lambda ~T_c^2 = \mu^2-m^2~,\quad
n = {1 \over 2} \mu T_c^2~.\label{two}
\eeq
These give a cubic equation for $X \equiv {m T_c^2 \over 2 n}$:
\beq
\nu X^3 = 1 - X^2~, \label{X}
\eeq
where $\nu \equiv {\lambda n \over 6 m^3}$ is a dimensionless constant
of order $N^0$.  Let $X_0 (\nu)$ be the solution that lies between $0$
and $1$.  (See Fig.~1.)  Then, we obtain
\beq
T_c^2 = {12 m^2 \over \lambda} \nu X_0 (\nu)~,\quad
\mu^2 - m^2 = m^2 \nu X_0 (\nu)~.
\eeq
The function $\nu X_0 (\nu)$ increases as $\nu$, and it behaves as
$\nu^{2 \over 3}$ for $\nu \gg 1$ and $\nu - {\nu^2 \over 2}$ for $\nu
\ll 1$.  For a very small coupling such that $\nu \ll 1$, we find
\beq
T_c^2 \simeq {2 n \over m} - \lambda~{n^2 \over 6 m^4}~.
\eeq
Therefore, the critical temperature decreases as the coupling increases.
This tendency also exists in the low temperature Bose condensation
phenomena.

\vskip .3cm
\centerline{\epsfxsize=\hsize \epsfbox{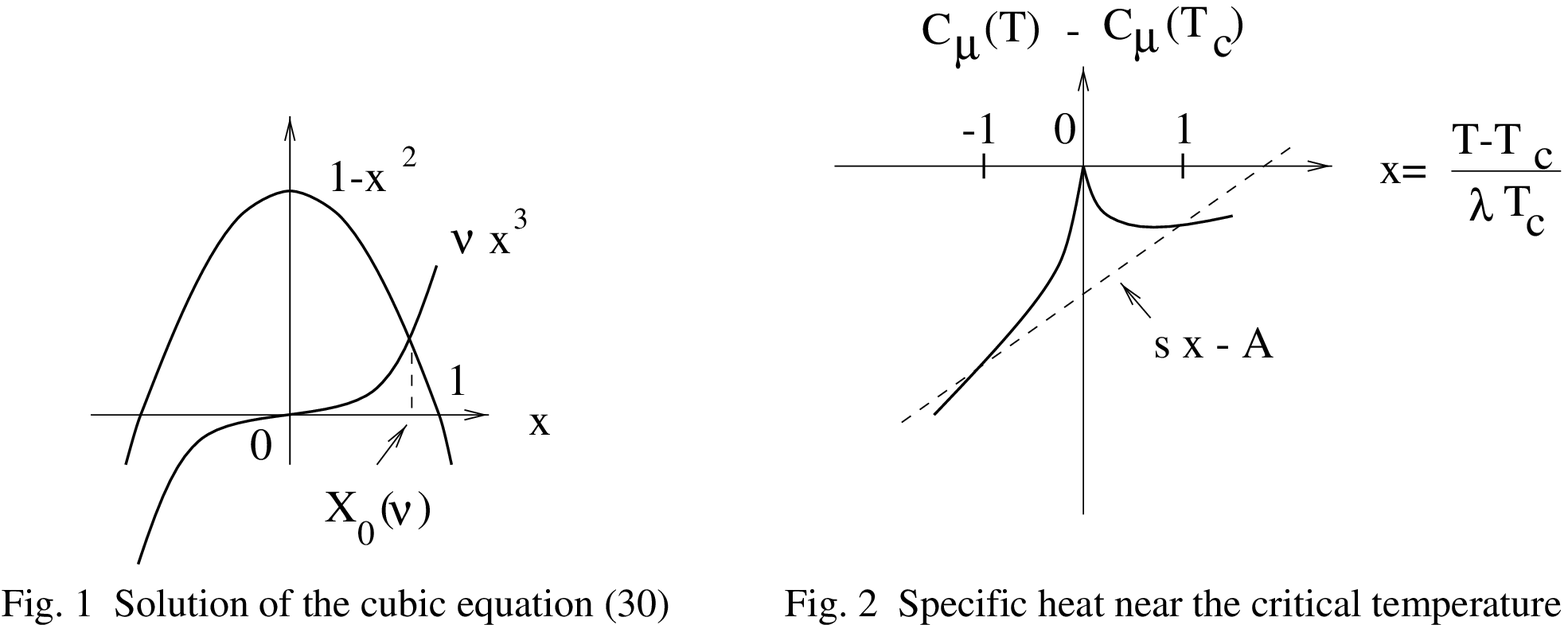}}

Finally, let us compute the temperature dependence of the specific heat
near the critical temperature $T_c$.  Using the free energy density of
the three dimensional effective theory given by
eqns.~(\ref{F3}--\ref{f3bar}), we obtain
\bea
C_\mu (T,\mu) &\equiv& - T {\p^2 Y_4 (T,\mu) \over (\p T)^2} = - T {\p^2
(T F_3) \over (\p T)^2} \nonumber\\ &\simeq& C_\mu (T_c,\mu) + s {T-T_c
\over \lambda T_c} - A \left| {T-T_c \over \lambda
T_c}\right|^{-\alpha}~,\label{Cmu}
\eea
where
\bea
C_\mu (T_c,\mu) &=& {(4 \pi)^2 \over 60}~T_c^3 + {1 \over 3} m^2 T_c~,\\ 
s &=& {(4\pi)^2 \over 20} \lambda T_c^3~,\quad
A = {1 \over 6^{3\over y_E}}
{3 \over y_E} \left({3 \over y_E} - 1\right) a~\lambda T_c^3~,
\eea
and
\beq
\alpha \equiv 2 - {3 \over y_E} \simeq - 0.01~.
\eeq
The above expression (\ref{Cmu}) is valid for $|T-T_c|/(\lambda T_c) =
{\rm O} (N^0)$, and we have ignored the terms of order $N \Lambda^3$.
By keeping only the first two terms of eqn.~(\ref{f3bar}), we have also
ignored $\left(|T-T_c|/(\lambda T_c)\right)^{- y'/y_E} \ll 1$.  The
positive constant $a$ in $A$ is the unknown constant in
eqn.~(\ref{f3bar}).  In Fig.~2 we show the specific heat schematically.
The specific heat at constant number density, $C_n$, differs from
$C_\mu$ only by a constant up to order $N^2 \Lambda^3$:
\beq
C_n - C_\mu \simeq - 2 \mu^2 T_c \simeq - 2 T_c \left( m^2 + {\lambda
\over 12}~T_c^2 \right)~.
\eeq

\vskip .2cm 

In conclusion we have shown how to apply the method of three dimensional
effective theory to understand Bose condensation at high temperatures.

\vskip .5cm

I would like to thank Prof.~K.~Kuboki for discussions.

\vfill\eject

\end{document}